\documentclass[conference,10pt]{IEEEtran}

\usepackage{amsmath}
\usepackage{amsthm}
\usepackage{epsfig}
\usepackage{cite}
\usepackage{amssymb}
\usepackage{psfrag}
\usepackage[latin1]{inputenc}
\usepackage{latexsym}
\usepackage[all]{xy}
\usepackage{subfigure}
\usepackage{color}
\newtheorem{theorem}{Theorem}

\title{Generalized Distributed Network Coding Based on Nonbinary Linear Block Codes for Multi-User Cooperative Communications}

\author{\IEEEauthorblockN{  João Luiz Rebelatto,
                            Bartolomeu F. Uchôa-Filho}
\IEEEauthorblockA{Communications Research Group\\
                    Department of Electrical Engineering \\
                    Federal University of Santa Catarina\\
                    Florianópolis, SC, 88040-900, Brazil \\
                    Email: \{jlrebelatto, uchoa\}@eel.ufsc.br}
\and
\IEEEauthorblockN{Yonghui Li and
                  Branka Vucetic}
\IEEEauthorblockA{Telecommunications Group \\
                    Department of Electrical Engineering\\
                    University of Sydney \\
                    NSW 2006, Australia \\
                    Email:  \{lyh, branka\}@ee.usyd.edu.au}}

\begin{document}
\maketitle
\begin{abstract}
In this work, we propose and analyze a generalized construction of distributed network codes for a network consisting of $M$ users sending different information to a common base station through independent block fading channels. The aim is to increase the diversity order of the system without reducing its code rate. The proposed scheme, called generalized dynamic-network codes (GDNC), is a generalization of the dynamic-network codes (DNC) recently proposed by Xiao and Skoglund. The design of the network codes that maximizes the diversity order is recognized as equivalent to the design of linear block codes over a nonbinary finite field under the Hamming metric. The proposed scheme offers a much better tradeoff between rate and diversity order. An outage probability analysis showing the improved performance is carried out, and computer simulations results are shown to agree with the analytical results.
\end{abstract}

\section{Introduction} \label{sec:introduction}

In a cooperative wireless communications system with multiple users transmitting independent information to a common base station (BS), besides broadcasting their own information, users help each other relaying their partners information~\cite{sendonaris.03,laneman.04,xiao.09,xiaoL.07}. In the \emph{decode-and-forward} (DF) relaying protocol~\cite{sendonaris.03,laneman.04}, the codeword relayed to the base station (BS) is a re-encoded version of the previously decoded codeword received in the broadcasting phase. 

Network coding~\cite{koeter.03}, a method originally proposed to attain maximum information flow in a network, has recently been applied in cooperative wireless communications systems to improve the bit error rate (BER) performance~\cite{xiao.09,xiao.09.ISIT,xiaoL.07}. In a network coded system, relays process information from different users and perform linear combinations of the received signals, with coefficients chosen from a finite field GF$(q)$.

A two-user cooperative system that employs binary network coding was proposed in~\cite{xiaoL.07}. In that scheme, each user transmits the binary sum (XOR) of its own source message and the received message from its partner (if correctly decoded). However, the scheme proposed in~\cite{xiaoL.07} does not improve the system diversity order.

In~\cite{xiao.09.ISIT}, it was shown that binary network coding are not optimal for achieving full diversity in multiple user-multiple relays system. A similar result was shown in~\cite{xiao.09}, however, instead of considering dedicated relays, therein the $M$ users themselves act as relays for each other. The scheme proposed in~\cite{xiao.09}, called \emph{dynamic-network codes} (DNC), considers a fixed nonbinary network code. In the first time slot, each user broadcasts a single packet of its own to the BS as well as to the other users, which try to decode the packet. From the second time slot until the $M$-th time slot, each user transmits to the BS $M-1$ nonbinary linear combinations of the packets that it could successfully decode. With DNC, by using an appropriately designed network code,  the diversity order was shown to be higher than in binary network coded systems. The scheme is called ``dynamic'' in the sense that the network code is designed to perform well under the possible occurrence of errors in the inter-user channels.

 In this paper, we elaborate on the DNC scheme by first recognizing the problem as equivalent to that of designing linear block codes over GF($q$) for erasure correction. In particular, for perfect inter-user channels, the diversity order equals the minimum Hamming distance of the block code, so the network transfer function should correspond to the generator matrix of an optimal block code under the Hamming metric. The Singleton bound appears as a natural upper bound on the diversity order, and this bound is achieved with a sufficiently large field size \cite{grassl.10}.

We then extend the DNC scheme by allowing each user to broadcast several (as opposed to just one) packets of its own in the broadcast phase, as well as to transmit several nonbinary linear combinations (of all correctly decoded packets) in the cooperative phase. From the block coding perspective, the so-called {\em generalized dynamic-network codes} (GDNC) consider a longer codeword, with more parity symbols, which improves the Singleton bound.

Yet from another point of view, in a GNDC system, temporal diversity is exploited, and we show that a much better tradeoff between rate and diversity order can be achieved, e.g., it is possible to set both the rate and the diversity order to be higher than that in the DNC scheme.

The rest of this paper is organized as follows. The next section presents the system model and some relevant previous works, including binary network coded cooperative systems~\cite{xiaoL.07} and the DNC scheme~\cite{xiao.09}. The motivation for GDNC is presented in Section~\ref{sec:motivation}. Section~\ref{sec:gdnc} presents the proposed GDNC scheme, showing firstly a simple 2-user network and then generalizing it to $M$ users. Simulations results are presented in Section~\ref{sec:simulations}. Finally, Section~\ref{sec:conclusions} presents our conclusions and final comments.

\section{Preliminaries} \label{sec:preliminaries}

\subsection{System Model}

The network consists of multiple users ($M \geq 2$) having different information to send to a common base station (BS).
One time slot (TS) is defined as the time period in which all the $M$ users realize a single transmission each one (through orthogonal channels, either in time, frequency or code), that is, one TS corresponds to $M$ transmissions.
The received baseband codeword at user $i$ at time $t$ is given by:
\begin{equation}
y_{j,i,t} = h_{j,i,t}x_{j,i,t} + n_{j,i,t},
\end{equation}
where $j \in \{1,\cdots,M\}$ represents the transmitter user and $i \in \{0,1,\cdots,M\}$ the receiver user (0 corresponds to the BS). The index $t$ denotes the time slot. $x_{j,i,t}$ and $y_{j,i,t}$ are transmitted and received codewords, respectively. $n_{j,i,t}$ is the zero-mean additive white Gaussian noise with variance $N_0/2$ per dimension. The channel gain due to multipath is denoted by $h_{j,i,t}$, as in \cite{xiao.09}, and it is assumed to have i.i.d. (across space and time) Rayleigh distribution with unit variance.

Assuming the $x_{j,i,t}$'s to be i.i.d. Gaussian random variables and considering all the channels with the same average  signal-to-noise ratio (SNR), the mutual information $I_{j,i,t}$ between $x_{j,i,t}$ and $y_{j,i,t}$ is:
\begin{equation} \label{eq:mutual_inf}
I_{j,i,t} = \log(1+|h_{j,i,t}|^2\text{SNR}).
\end{equation}

   Assuming powerful enough channel codes, $x_{j,i,t}$ can be correctly decoded if $I_{j,i,t}>R_{j,i,t}$, where $R_{j,i,t}$ is the information rate from user $j$ to user $i$ in the time slot $t$. Considering that all the users have the same rate, the index of $R$ can be dropped. Thus, $x_{j,i,t}$ cannot be correctly decoded if:
\begin{equation}
|h_{j,i,t}|^2<g,
\end{equation}
where $g=\frac{2^R-1}{\text{SNR}}$. The probability that such an event happens is called the \emph{outage probability}. For Rayleigh fading, the outage probability is calculated as~\cite{laneman.04,tse.05}:
\begin{equation}
P_e = \Pr\left\{|h_{j,i,t}|^2<g \right\}=1-e^{-g}\approx g.
\end{equation}

The approximation holds for high SNR region. Considering block fading, the diversity order $D$ is defined as~\cite{tse.05}:
\begin{equation} \label{eq:diversity}
D \triangleq \lim_{\text{SNR}\rightarrow \infty} \frac{-\log P_e}{\log \text{SNR}}.
\end{equation}

In this work, block fading means that fading coefficients are independent identically distributed (i.i.d.) random variables for different blocks but constant during the same block. It is also assumed that receivers have perfect channel state information (CSI), but the transmitters do not have any CSI.

\subsection{Binary Network Coded Cooperation} \label{subsec:2user}

Instead of transmitting only the partner's information in the second TS, as in the DF scheme, each user can transmit a binary sum of its own information and its partner's information. It was shown in~\cite{xiaoL.07} that the outage probability for a 2-user binary network coded system is given by:
\begin{equation} \label{eq:out_net_bin}
P_{o,\text{BNC}}\approx P_e^2,
\end{equation}
which corresponds to a diversity order $D=2$ according to \eqref{eq:diversity}. We can see that the diversity order obtained from \eqref{eq:out_net_bin} is not an optimal result, since the information of each user is transmitted through three independent paths and higher diversity order can be achieved, as will be explained later. By a similar analysis, the outage probability of the DF scheme is given by~\cite{laneman.04}:
\begin{equation}
P_{o,{\text{DF}}} \approx 0.5P_e^2.
\end{equation}

\subsection{Dynamic-Network Codes} \label{subsec:xiao}

\begin{figure}[!htb]
\center{
\hfill
\subfigure[Broadcast phase.\label{fig:xiaoa}]{
\xymatrix@C=1.5cm@-2pc{
\text{\footnotesize User 1} & & \\
\bigcirc \ar@{->}[rrdd]^-{I_1} \ar@{->}@<0.5ex>[ddd]^-{I_1}& &                 \\
& & \text{{\footnotesize BS}}\\
& &  \bigcirc\\
\bigcirc \ar@{->}[rru]_-{I_2}  \ar@{->}@<0.5ex>[uuu]^-{I_2}  & & \\
\text{\footnotesize User 2} & &   \\  \\
}}\hfill
\subfigure[Cooperative phase.\label{fig:xiaob}]{
\xymatrix@C=1.5cm@-2pc{
\text{\footnotesize User 1} & & \\
\bigcirc \ar@{->}[rrdd]^-{I_1+I_2}  & &                 \\
& & \text{{\footnotesize BS}}\\
& &  \bigcirc\\
\bigcirc \ar@{->}[rru]_-{I_1+2I_2}  & & \\
\text{\footnotesize User 2} & &          \\        \\
}}\hfill}
\caption{Two-user cooperative network with nonbinary network coding. (a) Each user broadcasts its own information and (b) each user transmits a linear combination on GF(4) of all the available information.}
\label{fig:Xiao}
\end{figure}
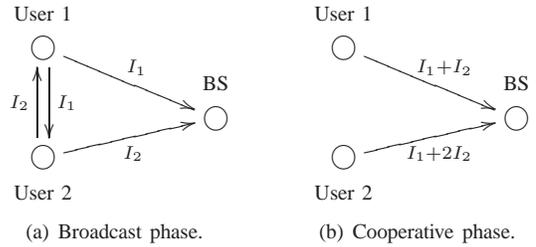

In \cite{xiao.09}, Xiao and Skoglund showed that the use of nonbinary network coding is necessary to achieve a higher diversity order and then proposed the so-called DNC. 
A simple 2-user DNC scheme is presented in Fig. \ref{fig:Xiao}, where nonbinary coefficients are now used.

Considering perfect inter-user channels, the BS receives $I_1$, $I_2$, $I_1\oplus I_2$ and $I_1\oplus 2I_2$. We can see that the BS is able to recover original messages $I_1$ and $I_2$ from any 2 out of 4 received codewords. For User 1 (the same result holds for user 2 due to symmetry), an outage occurs when the direct codeword $I_1$ and at least 2 out of the 3 remainder codewords cannot be decoded. This occurs with probability~\cite{xiao.09}
\begin{equation}
P_0=P_e\left[{{3}\choose{2}}P_e^2(1-P_e)+P_e^3\right] \approx 3P_e^3.
\end{equation}

However, with probability $P_e$ a user cannot decode its partner's codeword. In this situation, it retransmits its own information. The BS performs maximum ratio combining (MRC), which was shown in \cite{laneman.04} to have outage probability $P_1=P_e^2/2$. Thus, the overall outage probability for User 1 is~\cite{xiao.09}:
\begin{equation}
P_{o,1}= P_eP_1+(1-P_e)P_0 \approx 3.5P_e^3.
\end{equation}

It is easy to see that the diversity order is $D=3$. If the inter-user channels are not reciprocal, then the outage probability  was shown to be~\cite{xiao.09}:
\begin{equation} \label{eq:DNC_not_rec_M2}
P_{o,1}\approx 4P_e^3.
\end{equation}

For $M$ users, in the DNC scheme, each user transmits $M-1$ nonbinary linear combinations in the cooperative phase.  It is shown in \cite{xiao.09} that the diversity order achieved by DNC is $D=2M-1$, but with a fixed and low rate $R=M/M^2=1/M$.

\section{Motivation for this Work} \label{sec:motivation}

The multiple-sources-one-destination network can be represented by a transfer matrix which in turn can be seen as a generator matrix of a systematic linear block code. The generator matrix obtained from the 2-user system illustrated in Fig. \ref{fig:Xiao} is given by
\begin{equation} \label{eq:G}
\textbf{G}_{\text{DNC}} =
\left[
\begin{array}{cc|cc}
1 & 0 & 1 & 1 \\
0 & 1 & 1 & 2 \\
\end{array} \right].
\end{equation}

In the DNC scheme, the diversity order is related to the minimum number of correctly received packets (or symbols) at the BS with which the information packets from all users can be recovered. In terms of block coding, this is equivalent to the erasure correction capability of the block code. It is well-known that the transmitted codeword of a linear block code with minimum distance $D$ can be recovered if up to $D-1$ of its positions have been erasured by the channel~\cite{costello.04}. The connection between these two problems establishes that the diversity order of the 2-user DNC system, under the assumption of perfect inter-user channels, is equal to the minimum Hamming distance of the rate 2/4 block code with generator matrix given in \eqref{eq:G}.

In general, for a rate $k/n$ code, the minimum distance is upper bounded by the Singleton bound~\cite{costello.04}:
 \begin{equation} \label{eq:singleton}
 d_{\text{min}} \leq n-k+1.
 \end{equation}
 According to \cite{grassl.10}, the Singleton bound is achieved if the alphabet size is large enough. For example, for a $4/8$ block code, the Singleton bound gives $d_{\text{min}}\leq 5$. However, the maximum possible achieved minimum distance in GF$(2)$ is 3. In GF$(4)$, it is possible to achieve $d_{\text{min}}=4$. The upper bound $d_{\text{min}}=5$ is only possible to be achieved if the field size is at least 8.

In the DNC scheme, the overall rate is $M/M^2$. From \eqref{eq:singleton}, the diversity order is thus upper bounded by $D_{\max,\text{DNC}}=M^2-M+1$. A system with rate $R=\alpha M/ \alpha M^2$ (for $\alpha \geq 2$) would have the same overall rate as the DNC scheme ($R=1/M$), but the diversity upper bound would be increased to
\begin{equation} \label{eq:singleton_new}
D_{\max,\alpha}=\alpha (M^2-M)+1.
\end{equation}
This motivates us to modify the DNC scheme accordingly. In the next section, we propose an even more general scheme which is more flexible in terms of rate and diversity order.

Due to inter-user channel outages, the upper bound presented in \eqref{eq:singleton_new} cannot be achieved. The discrepancy between the upper bound and the real diversity order obtained is quantified by the outage probability analysis, in the next section.

\section{Generalized Dynamic-Network Codes} \label{sec:gdnc}

\begin{figure}[!htb]
\centering
\subfigure[Broadcast phase 1.\label{fig:newa}]{
\xymatrix@C=1.1cm@-2pc{
\text{\footnotesize User 1} & & \\
\bigcirc \ar@{->}[rrdd]^-{I_1(1)} \ar@{->}@<0.5ex>[ddd]^-{I_1(1)}& &                 \\
& & \text{{\footnotesize BS}}\\
& &  \bigcirc\\
\bigcirc \ar@{->}[rru]_-{I_2(1)}  \ar@{->}@<0.5ex>[uuu]^-{I_2(1)}  & & \\
\text{\footnotesize User 2} & &   \\  \\
}}
\subfigure[Broadcast phase 2.\label{fig:newb}]{
\xymatrix@C=1.1cm@-2pc{
\text{\footnotesize User 1} & & \\
\bigcirc \ar@{->}[rrdd]^-{I_1(2)} \ar@{->}@<0.5ex>[ddd]^-{I_1(2)}& &                 \\
& & \text{{\footnotesize BS}}\\
& &  \bigcirc\\
\bigcirc \ar@{->}[rru]_-{I_2(2)}  \ar@{->}@<0.5ex>[uuu]^-{I_2(2)}  & & \\
\text{\footnotesize User 2} & &   \\  \\
}}
\subfigure[Broadcast phase 3.\label{fig:newc}]{
\xymatrix@C=1.1cm@-2pc{
\text{\footnotesize User 1} & & \\
\bigcirc \ar@{->}[rrdd]^-{I_1(3)} \ar@{->}@<0.5ex>[ddd]^-{I_1(3)}& &                 \\
& & \text{{\footnotesize BS}}\\
& &  \bigcirc\\
\bigcirc \ar@{->}[rru]_-{I_2(3)}  \ar@{->}@<0.5ex>[uuu]^-{I_2(3)}  & & \\
\text{\footnotesize User 2} & &   \\  \\
}} \\
\subfigure[Cooperative phase 1.\label{fig:newd}]{
\hspace{0.5cm}
\xymatrix@C=1.1cm@-2pc{
\text{\footnotesize User 1} & & \\
\bigcirc \ar@{->}[rrdd]^-{\oplus_{1,1}}  & &                 \\
& & \text{{\footnotesize BS}}\\
& &  \bigcirc\\
\bigcirc \ar@{->}[rru]_-{\oplus_{2,1}}  & & \\
\text{\footnotesize User 2} & &          \\        \\
}
\hspace{0.5cm}}
\hspace{0.2cm}
\subfigure[Cooperative phase 2.\label{fig:newe}]{
\hspace{0.5cm}
\xymatrix@C=1.1cm@-2pc{
\text{\footnotesize User 1} & & \\
\bigcirc \ar@{->}[rrdd]^-{\oplus_{1,2}}  & &                 \\
& & \text{{\footnotesize BS}}\\
& &  \bigcirc\\
\bigcirc \ar@{->}[rru]_-{\oplus_{2,2}}  & & \\
\text{\footnotesize User 2} & &          \\        \\
}
\hspace{0.5cm}}
\caption{A rate 6/10 GDNC scheme with $M=2$ users. The symbol $\oplus_{m,t}$ denotes the linear combination of all the available information performed by relay $m$ in time slot $t$. For a 6/10 block code, according to \cite{grassl.10}, the minimum field size necessary to achieve $d_{\min}=5$ is $q=9$.}
\label{fig:New}
\end{figure}
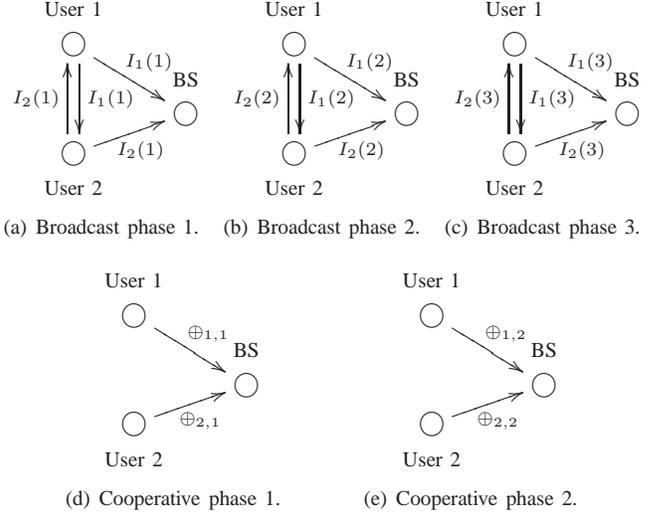

We begin the description of the proposed GDNC scheme by elaborating on the 2-user DNC scheme presented in Fig. \ref{fig:Xiao}, whose associated generator matrix is given in (\ref{eq:G}). The new scheme is illustrated in Fig. \ref{fig:New}.

Each user sends three packets (or symbols) of its own in the broadcast phase, and then each user sends two nonbinary linear combinations (of the six previously transmitted packets) over GF($q$) in the cooperative phase. The receiver collects the 10 packets, which can be seen as a codeword of a systematic 6/10 linear block code.

Without loss of generality, we analyze the outage probability for User 1 in the first TS. Consider for the moment being that the inter-user channels are reciprocal. In this case, the probability that no errors occur in the inter-user channel is $P_{p,1}=1-P_e$. If User 2 can correctly decode $I_1(1)$, the message $I_1(1)$ will suffer an outage at the BS when the direct packet \emph{and}, in the worst case, the 4 parity packets containing $I_1(1)$ cannot be decoded by the BS, which happens with probability $P_0 \approx P_e^5$.

If User 2 cannot correctly decode $I_1(1)$, which happens with probability $P_e$, it will not be able to help User 1 by relaying $I_1(1)$. In this case, the BS receives only 3 packets containing $I_1(1)$ (the direct transmission plus 2 parities sent by User 1 itself). Message $I_1(1)$ will be in outage at the BS if the direct packet \emph{and}, in the worst case, both of the parities transmitted by User 1 cannot be decoded. This happens with probability $P_1 \approx P_e^3$. Therefore, considering reciprocal inter-user channels and all the outage patterns, the outage probability of message $I_1(1)$ is given by:
\begin{equation} \label{eq:new_2user_outage}
P_{o,1} = P_e P_1+(1-P_e)P_0 \approx P_e^4.
\end{equation}
The same result is obtained when the inter-user channels are not reciprocal.

We can see from \eqref{eq:new_2user_outage} that the diversity order achieved by the rate 6/10 GDNC scheme with $M=2$ users presented in Fig. \ref{fig:New} is $D=4$, which is higher than the one obtained by the rate 2/4 DNC scheme with $M=2$ users in \eqref{eq:DNC_not_rec_M2}. Therefore, both the rate and the diversity order have been increased.

We can also verify that the outage probability is dominated by the term related to the inter-user channel being in outage, when User 2 cannot help User 1. Similarly, we will see that when all the $M-1$ inter-user channels fail is also the worst case scenario in the more general GDNC scheme for a $M$-user network.

\subsection{Multiple Users} \label{subsec:new_mult}

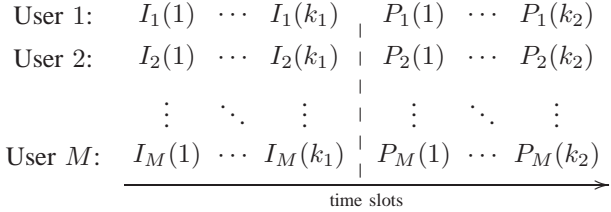
\begin{figure}[htb]
\[
\xymatrix@-2pc{
\txt{User $1$:}& & I_1(1)& \cdots  & I_1(k_1)&\ar@{--}[dddd] & P_1(1)& \cdots  & P_1(k_2)  \\
\txt{User $2$:}& & I_2(1)& \cdots  & I_2(k_1)& & P_2(1)& \cdots  & P_2(k_2) \\
& &\vdots & \ddots  & \vdots & &\vdots & \ddots  & \vdots \\
\txt{User $M$:}& & I_M(1)& \cdots & I_M(k_1)& & P_M(1)& \cdots & P_M(k_2)  \\
 & \ar[rrrrrrrr]_-{\text{time slots}} & & &&&&&&\\
}
\]
\caption{GDNC scheme for a network with $M$ users.}
\label{fig:Gen_proposed}
\end{figure}

The generalization of the scheme presented in Fig. \ref{fig:New} is shown in Fig. \ref{fig:Gen_proposed}. $I_j(t)$ represents the information transmitted by user $j$ ($j=1,\cdots,M$) in time slot $t$ ($t=1,\cdots,k_1$) of the broadcast phase (at left of the vertical dashed line), and $P_j(t')$ corresponds to the parity transmitted by user $j$ in time slot $t'$ ($t'=1,\cdots,k_2$) of the cooperative phase (at right of the vertical dashed line).

Following Fig. \ref{fig:Gen_proposed}, each user first broadcasts $k_1$ independent information packets\footnote{These packets can be sent in any order. Recall that the channels between two consecutive transmissions are considered uncorrelated across time and space.}. In the cooperative phase, each user transmits $k_2$ parity packets consisting if nonbinary linear combinations of its $k_1$ own information packets and the $k_1(M-1)$ partners' information packets (if decoded correctly). If an user cannot correctly decode an information packet from one of its partners, this information packet is replaced by an all-zero packet in the formation of the linear combination. Thus, the GDNC overall rate is given by:
\begin{equation} \label{eq:new_rate}
R_{\text{GDNC}} = \frac{k_1M}{k_1M+k_2M} = \frac{k_1}{k_1+k_2}.
\end{equation}
By varying $k_1$ and $k_2$ independently, we can have a tradeoff between rate and diversity. An appropriate choice of $k_1$ and $k_2$ can make the GDNC scheme to simultaneously improve rate and diversity order over the DNC scheme. When $k_1=1$ and $k_2=M-1$, the proposed scheme reduces to the DNC scheme. In particular, for $k_1 = k_2 = 1$ we have the 2-user DNC scheme in Fig. \ref{fig:Xiao}.

From (\ref{eq:singleton}) and (\ref{eq:new_rate}), we can see that the diversity order of the GDNC scheme is upper bounded by
\begin{equation} \label{eq:UB_GDNC}
D_{\text{max,GDNC}} \leq k_2M+1.
\end{equation}
As we already know, due to the errors in inter-users channels, this upper bound cannot be achieved. In the next section, we find the diversity order guaranteed by the proposed scheme.

\subsection{Outage Probability and Diversity Order} \label{subsec:gdnc_out}

Denote by $D_{j,t} \subseteq  \{ 1, \ldots, M \}$ the index set corresponding to the users that correctly decoded $I_j(t)$, the information packet of user $j$ in time slot $t$ in the broadcast phase. For convenience, include index $j$ itself to $D_{j,t}$. The number of users in $D_{j,t}$ is denoted by $|D_{j,t}|$. Let $P_e$ again be the outage probability of a single channel, and let $\overline{D}_{j,t}$ denote the complement set $\{ 1, \ldots, M \} \backslash D_{j,t}$, {\em i.e.}, $\overline{D}_{j,t}$ contains the indices of the users which could not decode $I_j(t)$ correctly. The probability of $\overline{D}_{j,t}$ is approximately $P_e^{|\overline{D}_{j,t}|}$. We should note that the message $I_j(t)$ is contained in $(M-|\overline{D}_{j,t}|)k_2+1$ packets (1 in the systematic part plus $(M-|\overline{D}_{j,t}|)k_2$ as part of parities) transmitted to the BS through independent channels. For a fixed $D_{j,t}$ (which fixes $\overline{D}_{j,t}$ as well), it can be shown that the probability that the BS cannot recover $I_j(t)$ is \[ P_{o,j}(\overline{D}_{j,t}) \approx  \gamma(k_1,k_2,\overline{D}_{j,t}) P_e^{(M-|\overline{D}_{j,t}|)k_2+1}, \]
where $\gamma(k_1,k_2,\overline{D}_{j,t})$ is a positive integer representing the number (multiplicity) of outage patterns leading to that same probability. In particular, $\gamma(k_1,k_2,\{1,\ldots,M \} \backslash \{j \}) = 1$.

The overall outage probability is then given by:
\begin{eqnarray} \label{eq:over_prob}
P_{o,j} & = & \sum_{\overline{D}_{j,t}} P_e^{|\overline{D}_{j,t}|}(1-P_e)^{(M-1)-|\overline{D}_{j,t}|} P_{o,j}(\overline{D}_{j,t}) \nonumber \\
& \approx & \sum_{\overline{D}_{j,t}}  P_e^{(M-|\overline{D}_{j,t}|)k_2+|\overline{D}_{j,t}|+1} \gamma(k_1,k_2,\overline{D}_{j,t}) \label{eq:newproba}\\
& \approx & {{M-1}\choose{|\overline{D}_{j,t}|^*}} P_e^{(M-|\overline{D}_{j,t}|^*)k_2+|\overline{D}_{j,t}|^*+1} \label{eq:newprobb} \\
\nonumber \\
& = & P_e^{M+k_2}, \label{eq:newprobc}
\end{eqnarray}
where $P_e^{|\overline{D}_{j,t}|}(1-P_e)^{(M-1)-|\overline{D}_{j,t}|}$ is the probability of $|\overline{D}_{j,t}|$ out of $M-1$ inter-user channels in time slot $t$ being in outage, and $|\overline{D}_{j,t}|^*$ corresponds to the $|\overline{D}_{j,t}|$ value that results in the lowest exponent term in \eqref{eq:newproba}, which, for $k_2 \ge 2$, is $|\overline{D}_{j,t}|^*=M-1$. In (\ref{eq:newprobb}), ${{n}\choose{k}}$ is the binomial coefficient. We have proved the following result.

\begin{theorem} \label{th:new_diversity}
The diversity order of the GDNC scheme for sufficiently large field size is $D_{\text{GDNC}}=M+k_2$.
\end{theorem}
 We can see that, when $k_1=1$ and $k_2=M-1$, the proposed scheme reduces to the DNC scheme, with rate $1/M$ and diversity order $2M-1$, as mentioned at the end of Section \ref{subsec:xiao}.

\section{Simulation Results} \label{sec:simulations}

In order to support the results obtained through the outage probability analysis of the previous section, we have performed some computer simulations. The frame error rate (FER) was simulated and plotted against the SNR. The analytical outage probabilities obtained in this paper were also plotted (dashed lines). We assume that there exists a channel code with which it is possible to recover the transmitted packet if $|h_{j,i,t}|^2\geq g$. If $|h_{j,i,t}|^2 < g$, an outage is declared. To construct the network code over GF($q$), we considered the generator matrix of a block code over GF($q$) with the largest possible minimum Hamming distance (upper bound on the diversity order as presented in \eqref{eq:UB_GDNC}).

\begin{figure} [!hbt]
\begin{center}
\resizebox{9cm}{!}{
\includegraphics{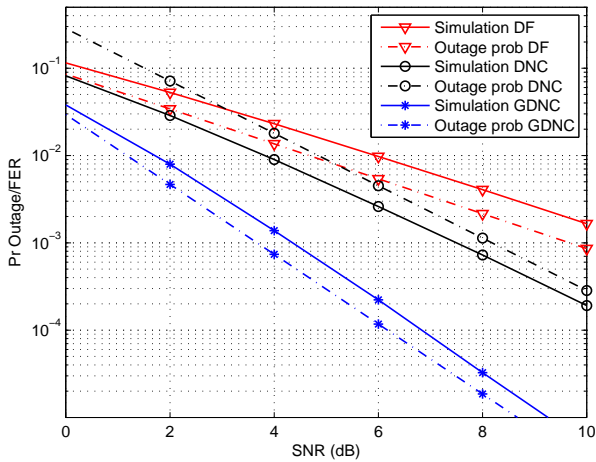}}\end{center}
\vspace{-0.5cm}
\caption{FER versus SNR (dB) for a 2-user system with rate $R=1/2$, considering the DF scheme, DNC scheme (in GF(4), according to \eqref{eq:G}) and the proposed GDNC scheme (with $k_1=k_2=2$ and in GF(8), according to \eqref{eq:G_GNDC}).}
\label{fig:FER1}
\end{figure}

Fig. \ref{fig:FER1} presents the FER versus SNR for a 2-user network, considering the DF scheme, the DNC scheme (Fig. \ref{fig:Xiao}) over GF(4), and the proposed GDNC with $k_1=k_2=2$ over GF(8), all of them with the same rate equal to 1/2. The generator matrix for the proposed system was chosen as:

\begin{equation} \label{eq:G_GNDC}
\textbf{G}_{\text{GDNC}} =
\left[
\begin{array}{cccc|cccc}
1 & 0 & 0 & 0 & 3 & 7 & 3&6 \\
0 & 1 & 0 & 0 & 5&7 &7 &4 \\
0 & 0 & 1 & 0 & 2& 4& 6& 1\\
0 & 0 & 0 & 1 &5 &5 &3 &2 \\
\end{array} \right].
\end{equation}

It should be mentioned that, since the maximum diversity order 3 of the DNC scheme is already achieved with the field GF(4), increasing the field size would not bring any advantages in this case. On the other hand, as discussed in Section \ref{sec:motivation}, for the GDNC used in our simulations, which corresponds to a rate 4/8 block code, the field size 8 was necessary for having $d_{\min}=5$.

As expected, the proposed scheme outperforms the two other schemes with the same rate.

 The SNR gap between the analytical and simulated results occurs due to the Gaussian input assumption made in \eqref{eq:mutual_inf}. However, we can see that the diversity order (curve slope) obtained by the simulations matches the one obtained analytically.


%

\section{Conclusions and Final Comments} \label{sec:conclusions}

In this work, we have proposed a generalization of the distributed network coding method in \cite{xiao.09}. The aim of the generalized dynamic-network code (GDNC) is to increase the diversity order of cooperative wireless communications systems without sacrifice in the system's rate, what appears to be a drawback of the original method. We have shown that the problem of designing the network codes that maximize the diversity order is related to that of designing optimal (in the Hamming sense) linear block codes over a nonbinary finite field.

An outage probability analysis was presented, and computer simulation results supported the analytical results.

The two design parameters $k_1$ and $k_2$ of GDNC may be varied to produce a wide range of rates and diversity orders, and offer a much better tradeoff between rate and diversity order (when the value of $k_2$ is changed), or even between rate and decoding latency (when $k_1$ is changed), as compared to the baseline system.

For a large number of users and/or when the parameters $k_1$ and $k_2$ are very large, the dimensions of the generator matrix of the block code associated with the GDNC scheme may be too large to be coped with by the encoders and decoders. In this case, a sparse generator matrix should be considered and the overall system could be seen as a generalization of the adaptive network coded cooperation (ANCC) method proposed in \cite{bao.08}. The connection between GDNC and ANCC is currently being investigated.

\section*{Acknowledgement} This work has been supported in part by CNPq (Brazil).


\bibliography{IEEEabrv,biblio}

\begin{thebibliography}{10}
\providecommand{\url}[1]{#1}
\csname url@samestyle\endcsname
\providecommand{\newblock}{\relax}
\providecommand{\bibinfo}[2]{#2}
\providecommand{\BIBentrySTDinterwordspacing}{\spaceskip=0pt\relax}
\providecommand{\BIBentryALTinterwordstretchfactor}{4}
\providecommand{\BIBentryALTinterwordspacing}{\spaceskip=\fontdimen2\font plus
\BIBentryALTinterwordstretchfactor\fontdimen3\font minus
  \fontdimen4\font\relax}
\providecommand{\BIBforeignlanguage}[2]{{%
\expandafter\ifx\csname l@#1\endcsname\relax
\typeout{** WARNING: IEEEtran.bst: No hyphenation pattern has been}%
\typeout{** loaded for the language `#1'. Using the pattern for}%
\typeout{** the default language instead.}%
\else
\language=\csname l@#1\endcsname
\fi
#2}}
\providecommand{\BIBdecl}{\relax}
\BIBdecl

\bibitem{sendonaris.03}
A.~Sendonaris, E.~Erkip, and B.~Aazhang, ``User cooperation diversity: {P}art
  {I} and {P}art {II},'' \emph{{IEEE} Trans. Commun.}, vol.~51, no.~11, pp.
  1927--1948, November 2003.

\bibitem{laneman.04}
J.~N. Laneman, D.~N.~C. Tse, and G.~W. Wornell, ``Cooperative diversity in
  wireless networks: Efficient protocols and outage bahavior,'' \emph{{IEEE}
  Trans. Inf. Theory}, vol.~50, no.~12, pp. 3062--3080, December 2004.

\bibitem{xiao.09}
M.~Xiao and M.~Skoglund, ``M-user cooperative wireless communications based on
  nonbinary network codes,'' in \emph{{Proc. IEEE Inform. Theory Workshop.
  ITW'09}}, June 2009, pp. 316 -- 320.

\bibitem{xiaoL.07}
L.~Xiao, T.~Fuja, J.~Kliewer, and D.~Costello, ``A network coding approach to
  cooperative diversity,'' \emph{{IEEE} Trans. Inf. Theory}, vol.~53, no.~10,
  pp. 3714--3722, October 2007.

\bibitem{koeter.03}
R.~Koetter and M.~Medard, ``An algebraic approach to network coding,''
  \emph{{IEEE/ACM} Trans. Netw.}, vol.~11, no.~5, pp. 782-- 795, October 2003.

\bibitem{xiao.09.ISIT}
M.~Xiao and M.~Skoglund, ``Design of network codes for multiple-user
  multiple-relay wireless networks,'' in \emph{{Proc. IEEE Intern. Symp. on
  Inform. Theory. ISIT'09}}, June 2009, pp. 2562 -- 2566.

\bibitem{grassl.10}
M.~Grassl, ``Bounds on the minimum distance of linear codes and quantum
  codes,'' Online available at http://www.codetables.de, Accessed on
  2010-01-02.

\bibitem{tse.05}
D.~Tse and P.~Viswanath, \emph{Fundamentals of Wireless Communications}.\hskip
  1em plus 0.5em minus 0.4em\relax Cambridge: Cambridge University Press, 2005.

\bibitem{costello.04}
S.~Lin and D.~J. Costello, \emph{Error Control Coding}, 2nd~ed.\hskip 1em plus
  0.5em minus 0.4em\relax Upper Saddle River, NJ, USA: Prentice Hall, 2004.

\bibitem{bao.08}
X.~Bao and J.~Li, ``Adaptive network coded cooperation ({ANCC}) for wireless
  relay networks: Matching code-on-graph with network-on-graph,'' \emph{{IEEE}
  Trans. Wireless Commun.}, vol.~7, no.~2, pp. 574--583, February 2008.

\end{thebibliography}
\bibliographystyle{ieeetran}

\end{document}